\documentclass[aps,twocolumn,pra,showpacs,tightenlines]{revtex4-1}
\usepackage{amsmath}
\usepackage{graphicx}
\usepackage{color}
\usepackage{amsfonts}
\usepackage[colorlinks,citecolor=blue]{hyperref}
\begin{document}

\title{Accelerated ground-state cooling of an optomechanical resonator via shortcuts \\to adiabaticity}
\author{Yu-Hong Liu}
\affiliation{Key Laboratory of Low-Dimensional Quantum Structures and Quantum Control of
Ministry of Education, Key Laboratory for Matter Microstructure and Function of Hunan Province, Department of Physics and Synergetic Innovation Center for Quantum Effects and Applications, Hunan Normal University, Changsha 410081, China}

\author{Xian-Li Yin}
\affiliation{Key Laboratory of Low-Dimensional Quantum Structures and Quantum Control of
Ministry of Education, Key Laboratory for Matter Microstructure and Function of Hunan Province, Department of Physics and Synergetic Innovation Center for Quantum Effects and Applications, Hunan Normal University, Changsha 410081, China}

\author{Jin-Feng Huang}
\email{Corresponding author: jfhuang@hunnu.edu.cn}
\affiliation{Key Laboratory of Low-Dimensional Quantum Structures and Quantum Control of
Ministry of Education, Key Laboratory for Matter Microstructure and Function of Hunan Province, Department of Physics and Synergetic Innovation Center for Quantum Effects and Applications, Hunan Normal University, Changsha 410081, China}

\author{Jie-Qiao Liao}
\email{Corresponding author: jqliao@hunnu.edu.cn}
\affiliation{Key Laboratory of Low-Dimensional Quantum Structures and Quantum Control of
Ministry of Education, Key Laboratory for Matter Microstructure and Function of Hunan Province, Department of Physics and Synergetic Innovation Center for Quantum Effects and Applications, Hunan Normal University, Changsha 410081, China}

\begin{abstract}
Ground-state cooling of mechanical resonators is an important task in quantum optomechanics, because it is a necessary prerequisite for the creation, manipulation, and application of macroscopic mechanical coherence. Here, we propose a transient-state scheme to accelerate ground-state cooling of a mechanical resonator in a three-mode loop-coupled optomechanical system via shortcuts to adiabaticity (STA). We consider four kinds of coupling protocols and calculate the evolution of the mean phonon number of the mechanical resonator in both the adiabatic and STA cases. We verify that the ground-state cooling of the mechanical resonator can be achieved with the STA method in a much shorter period. The STA method can be generalized to accelerate other adiabatic processes in cavity optomechanics, and hence this work will open up a realm of fast optomechanical manipulations.
\end{abstract}
\date{\today}
\maketitle

\section{INTRODUCTION}

Ground-state cooling of mechanical resonators (MRs) in cavity optomechanics has attracted great interest from both theoreticians and experimentalists in the fields of quantum optics and micro- and nanoscale physics~\cite{Vahala2008,Schwab2012,Kippenberg2014}. This is because the preparation of MRs into their ground states is a crucial step for the study of the fundamental of quantum mechanics~\cite{Schwab2005} such as macroscopic mechanical coherence and quantum decoherence~\cite{Zurek1991}, and the applications of optomechanical technologies~\cite{Metcalfe2014} such as quantum precise measurement~\cite{Schwab2004}. Up to now, ground-state cooling of a single MR in optomechanical systems~\cite{LiuChin2013} has been achieved mostly through two cooling methods: sideband cooling~\cite{Wilson-Rae2007PRL,Marquardt2007PRL,Dobrindt2008PRL,Genes2008NJP,Yong2008PRB,Chan2011Nature,Teufel2011Nature,Liu2013PRL1,He2017PRL,Xu2017PRL,Clarkl2017Nature,Qiu2020PRL} and feedback cooling~\cite{Rossi2017PRL,Rossi2018Nature,Conangla2019PRL,Tebbenjohanns2019PRL,Sommer2019PRL,Guo2019PRL,Sommer2020PRR}. These two methods generally require the systems to reach their steady states. Meanwhile, some transient-state-cooling schemes~\cite{Liu2013PRL1,Jacobs2011,Retzker2012,Yong2011,Liao2011,Chen2015,Sarma2020NJP} have been proposed in optomechanical systems. These schemes mainly introduce the modulation of system parameters, such as cavity dissipation~\cite{Liu2013PRL1}, input laser intensity~\cite{Jacobs2011,Retzker2012}, mechanical resonance frequency~\cite{Yong2011}, coupling strength~\cite{Liao2011,Chen2015}, and cavity frequency~\cite{Sarma2020NJP}. In particular, some of these transient-state-cooling schemes are based on the adiabatic evolutions~\cite{Sarma2020NJP}, which require slow evolution to satisfy the adiabatic condition. Note that the coherent excitation transfer between two mechanical modes in multimode optomechanical systems with the stimulated Raman adiabatic passage (STIRAP)~\cite{Gaubatz1990,Vitanov2017} method has been demonstrated in a recent experiment~\cite{Fedoseev2021}.

For adiabatically evolving systems, though their evolutions are robust to the parameter imperfections, they should evolve slowly to suppress the nonadiabatic transitions, which will accumulate decoherence in a practical evolution. Both the nonadiabatic transitions and the environmental decoherence usually lead to evolution error and low fidelity~\cite{Vitanov2017,Muga2019,Torrontegui2013}. Meanwhile, from the viewpoint of quantum operations, it is expected to implement fast quantum manipulations such that more operations can be completed in the coherence-preserved duration. In terms of cooling, how to realize a fast ground-state cooling of the MR in optomechanical systems becomes an interesting project.

To address this concern, we generalize the physical idea of the so-called shortcuts to adiabaticity (STA)~\cite{Muga2019,Torrontegui2013,Chen2010PRL1,Demirplak2003,Berry2009,Chen2010,Ibanez2012,Campo2013,Garaot2014,Baksic2016}
method to accelerate the cooling process but keep the merits of the adiabatic passage. The STA method constructs an explicitly auxiliary Hamiltonian $\hat{H}_{cd}(t)$ to eliminate nonadiabatic transitions and compel the system to follow the eigenstates of $\hat{H}_{\textrm{app}}(t)$~\cite{Berry2009,Chen2010}, thus implementing perfect excitation transfer at finite evolution periods~\cite{Deng2019,Xue2017,Xia2019}. In particular, the STA method has been experimentally implemented in various platforms~\cite{Suter2013,An2016,Du2016,Awschalom2017,Yin2017,Ness2018,GuoNJP2018,Guo2018,Long2018,Guo2019,Danilin2019,Maletinsky2019,Yu2019,Peng2020}, including nitrogen-vacancy-center systems~\cite{Suter2013,Awschalom2017,Maletinsky2019}, trapped ions~\cite{An2016,Guo2018}, cold-atom systems~\cite{Du2016,Ness2018}, superconducting circuits~\cite{Yin2017,GuoNJP2018,Guo2019,Danilin2019,Yu2019}, and nuclear-magnetic-resonance systems~\cite{Long2018,Peng2020}. In our work, we first use the STIRAP method to realize the ground-state cooling of the MR, which takes a long pulsed driving time. Then, based on the mapping relation between a three-mode system and a three-level system, we obtain the expected form of the auxiliary Hamiltonian $\hat{H}_{cd}(t)$ and study the cooling efficiency in the STA scheme. Compared with the STIRAP scheme, the STA method can not only achieve the ground-state cooling of a MR, but also increase the cooling velocities by two orders of magnitude. Additionally, the amplitudes of the pulsed driving fields can be accurately calculated. Accelerating the ground-state cooling of a MR with the STA method will inspire us with new ideas to accelerate other adiabatic processes in optomechanical systems, and provide new means for fast optomechanical manipulations.

The rest of this paper is organized as follows. In Sec.~\ref{sec:model}, we introduce the physical system and present the Hamiltonians. In Sec.~\ref {sec:adiabatic}, we consider the adiabatic cooling of the MR under four kinds of STIRAP protocols. In Sec.~\ref{sec:comparison}, we compare the differences and draw some conclusions among the four kinds of coupling protocols. In Sec.~\ref{sec:STA}, we study how to accelerate the ground-state cooling of a MR via STA. In Sec.~\ref{sec:EXPERIMENT}, we present some discussions on the experimental implementation of our scheme. Finally, we conclude this work in Sec.~\ref{sec:conclusion}. An appendix is presented to show the equations of motion for all the second-order moments.

\section{SYSTEM AND HAMILTONIAN\label{sec:model}}

\begin{figure}[tbp]
\centering \includegraphics[width=0.48\textwidth]{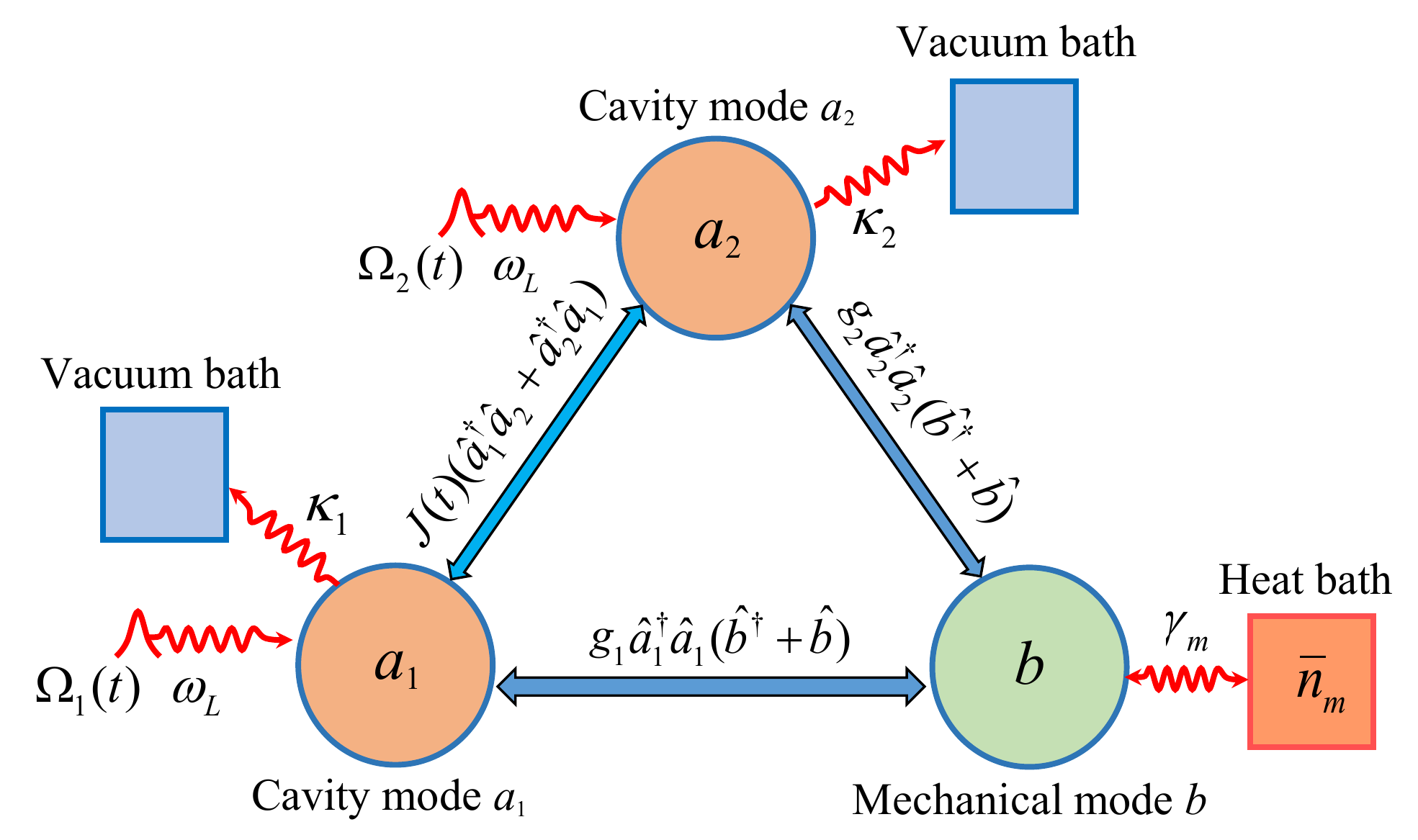}
\caption{(Color online) Schematic diagram of the three-mode loop-coupled optomechanical system, where the mechanical mode $b$ is optomechanically coupled to two cavity modes $a_{1}$ and $a_{2}$ with resonance frequencies $\omega_{1}$ and $\omega_{2}$, respectively. The two cavity modes are coupled with each other via a time-dependent photon-hopping interaction with coupling strength $J(t)$. To manipulate this system, the two cavity modes are driven by two pulsed fields with the carrier frequency $\omega_{L}$ and individual time-dependent driving amplitudes $\Omega _{1}\left( t\right) $ and $\Omega _{2}\left( t\right) $.}
\label{Fig1}
\end{figure}

We consider a three-mode optomechanical system (Fig.~\ref{Fig1}) that consists of a MR optomechanically coupled to two cavity-field modes, which are coupled with each other via a time-dependent photon-hopping interaction. In addition, the two cavity modes are driven by the respective pulsed fields. In a rotating frame defined by the unitary operator $\exp[-i\omega _{L}t(\hat{a}_{1}^{\dag }\hat{a}_{1}+\hat{a}_{2}^{\dag }\hat{a}_{2}) ]$, the Hamiltonian of the system is given by ($\hbar=1$)
\begin{eqnarray}
\hat{H}_{R}(t)  &=&\omega _{m}\hat{b}^{\dagger }\hat{b}+\sum_{i=1,2}[\Delta _{i}\hat{a}_{i}^{\dag }\hat{a}_{i}-g_{i}\hat{a}_{i}^{\dag }\hat{a}_{i}(\hat{b}^{\dagger }+\hat{b})]
\nonumber \\
&&\!+J(t) (\hat{a}_{1}^{\dag }\hat{a}_{2}+\hat{a}
_{2}^{\dag }\hat{a}_{1}) \!+\!\!\sum_{i=1,2}[ \Omega _{i}(t)\hat{a}_{i}^{\dag }\!+
\mathrm{H.c.}],\label{eq1}
\end{eqnarray}
where $\Delta _{i}=\omega_{i}-\omega _{L}$ (for $i=1,2$) is the driving detuning of the cavity-mode resonance frequency $\omega_{i}$ with respect to the carrier frequency $\omega_{L}$ of the driving pulse. The bosonic operators $\hat{a}_{i=1,2}$ $(\hat{a}_{i}^{\dagger })$ and $\hat{b}$ $(\hat{b}^{\dagger })$ are, respectively, the annihilation (creation) operators of the $i$th cavity mode $a_{i}$ and the mechanical mode $b$, with the corresponding resonance frequencies $\omega _{i}$ and $\omega _{m}$. The $g_{i=1,2}$ term in Eq.~(\ref{eq1}) describes the optomechanical coupling between the cavity mode $a_{i}$ and the mechanical mode $b$, with $g_{i}$ being the single-photon optomechanical-coupling strength. The $J(t)$ term denotes the photon-hopping coupling between the two cavity modes. The $\Omega _{i}(t)$ is the time-dependent driving amplitude associated with the pulsed driving field of the $i$th cavity mode.

In the open-system case, we assume that the two cavity modes are coupled to individual vacuum baths, and that the MR is coupled to a heat bath. In the Markovian-dissipation case, the evolution of the system is governed by the quantum master equation
\begin{eqnarray}
\dot{\hat{\rho}} &=&i[\hat{\rho},\hat{H}_{R}(t)]+\kappa _{1}%
\mathcal{\hat{D}}[\hat{a}_{1}]\hat{\rho}+\kappa _{2}\mathcal{\hat{D}}[\hat{a}%
_{2}]\hat{\rho}  \notag \\
&&+\gamma _{m}( \bar{n}_{m}+1) \mathcal{\hat{D}}[ \hat{b}]  \hat{
\rho}+\gamma _{m} \bar{n}_{m}\mathcal{\hat{D}}[\hat{b}^{\dag }]  \hat{
\rho},  \label{Eq2}
\end{eqnarray}%
where $\hat{\rho}$ is the density matrix of the three-mode system, $\hat{H}_{R}(t)$ is given by Eq.~(\ref{eq1}), and $\mathcal{\hat{D}}[\hat{o}]\hat{\rho}=\hat{o}\hat{\rho}\hat{o}^{\dag }-(\hat{o}^{\dag }\hat{o}\hat{\rho}+\hat{\rho}\hat{o}^{\dag }\hat{o})/2$ (for $\hat{o}=\hat{a}_{i=1,2}$, $\hat{b}$, and $\hat{b}^{\dag }$) is the standard Lindblad superoperator~\cite{Agarwal2013}. The parameters $\kappa_{i=1,2}$ and $\gamma _{m}$ are the decay rates of the $i$th cavity mode and the MR, respectively, and $\bar{n}_{m}$ is the environment thermal-excitation occupation of the MR.

To perform the linearization procedure, we make the displacement transformation to the quantum master equation~(\ref{Eq2}) by introducing the density matrix $\hat{\rho}'(t)$ in the displaced representation as
\begin{equation}
\hat{\rho}' = \hat{D}_{2}(\alpha _{2})\hat{D}_{1}(\alpha _{1})\hat{D}_{b}(\beta)\hat{\rho} \hat{D}_{b}^{\dag }(\beta ) \hat{D}_{1}^{\dag }(\alpha _{1})\hat{D}_{2}^{\dag }(\alpha _{2}),
\end{equation}
where $\hat{D}_{i}(\alpha _{i})=\exp (\alpha _{i}\hat{a}_{i}^{\dagger}-\alpha _{i}^{\ast }\hat{a}_{i})$ (for $i=1,2$) and $\hat{D}_{b}(\beta)=\exp(\beta \hat{b}^{\dag }-\beta ^{\ast }\hat{b})$ are the displacement operators, with the time-dependent displacement amplitudes $\alpha _{i}(t)$ and $\beta(t)$, respectively. In the displacement representation, the quantum master equation~(\ref{Eq2}) becomes
\begin{eqnarray}
\dot{\hat{\rho}}^{\prime}  &=&i[\hat{\rho}^{\prime},\hat{H}_{D}(t)]+\kappa _{1}%
\mathcal{\hat{D}}[\hat{a}_{1}]\hat{\rho}^{\prime}+\kappa _{2}\mathcal{\hat{D}}[\hat{a}%
_{2}]\hat{\rho}^{\prime}  \notag \\
&&+\gamma _{m}( \bar{n}_{m}+1) \mathcal{\hat{D}}[ \hat{b}]  \hat{
\rho}^{\prime}+\gamma _{m} \bar{n}_{m}\mathcal{\hat{D}}[\hat{b}^{\dag }]  \hat{
\rho}^{\prime},  \label{Eq4}
\end{eqnarray}%
where the displaced Hamiltonian is given by
\begin{eqnarray}
\hat{H}_{D}(t)  &=&\sum_{i=1,2}[\Delta _{i}+2g_{i}\mathrm{Re}[
\beta(t)] -g_{i}( \hat{b}^{\dagger }+\hat{b}) %
] \hat{a}_{i}^{\dag }\hat{a}_{i}  \nonumber \\
&&+\omega _{m}\hat{b}^{\dagger }\hat{b}+J( t) ( \hat{a}%
_{1}^{\dag }\hat{a}_{2}+\hat{a}_{2}^{\dag }\hat{a}_{1})   \nonumber \\
&&+\sum_{i=1,2}[ g_{i}\alpha _{i}(t)\hat{a}_{i}^{\dagger }( \hat{b}%
^{\dagger }+\hat{b}) +\mathrm{H.c.}]\label{A6}.
\end{eqnarray}
Here, $\mathrm{Re}[\beta(t)]$ gives the real part of $\beta(t)$. In Eq.~(\ref{A6}), the displacement amplitudes $\alpha_{i=1,2}(t)$ and $\beta(t)$ are determined by the following equations of motion
\begin{subequations}
\begin{eqnarray}
\dot{\alpha}_{1}&=&\left( -i\Delta _{1}-2ig_{1}\mathrm{Re}[\beta
] -\frac{\kappa _{1}}{2}\right) \alpha _{1}-iJ(t) \alpha
_{2}+i\Omega _{1}(t),\   \notag  \label{5} \\
&& \\
\dot{\alpha}_{2} &=&\left(-i\Delta _{2}-2ig_{2}\mathrm{Re}[\beta] -\frac{\kappa _{2}}{2}\right) \alpha _{2}-iJ( t) \alpha
_{1}+i\Omega _{2}(t),\   \notag \\
&& \\
\dot{\beta} &=&\left( -i\omega _{m}-\frac{\gamma _{m}}{2}\right) \beta
-ig_{1}|\alpha _{1}|^{2}-ig_{2}|\alpha _{2}|^{2}.
\label{6}
\end{eqnarray}
\end{subequations}
Below, we neglect the high-order interaction terms and consider the following parameter condition
\begin{eqnarray}
\Delta _{i=1,2}\gg 2g_{i}\mathrm{Re}[\beta],
\end{eqnarray}
Then, Hamiltonian (\ref{A6}) can be approximated as
\begin{eqnarray}
\hat{H}_{\textrm{app}}(t) &=&\sum_{i=1,2}\Delta _{i}\hat{a}_{i}^{\dagger }\hat{a}%
_{i}+\omega _{m}\hat{b}^{\dagger }\hat{b}+J(t)(\hat{a}_{1}^{\dagger }%
\hat{a}_{2}+\hat{a}_{2}^{\dagger }\hat{a}_{1})  \notag \\
&&+\sum_{i=1,2}[ G_{i}(t)\hat{a}_{i}^{\dagger }(\hat{b}^{\dagger }+\hat{b}%
)+\mathrm{H.c.}]\label{eqA8} ,
\end{eqnarray}
where $G_{i=1,2}\left( t\right) \equiv g_{i}\alpha _{i}(t)$ is the strength of the linearized optomechanical coupling between cavity mode $a_{i}$ and mechanical mode $b$.

Based on the above discussions, we know that the evolution of the linearized three-mode optomechanical system is governed by the quantum master equation
\begin{eqnarray}
\dot{\hat{\rho}}^{\prime}  &=&i[\hat{\rho}^{\prime},\hat{H}_{\textrm{app}}(t)]+\kappa _{1}%
\mathcal{\hat{D}}[\hat{a}_{1}]\hat{\rho}^{\prime}+\kappa _{2}\mathcal{\hat{D}}[\hat{a}%
_{2}]\hat{\rho}^{\prime}  \notag \\
&&+\gamma _{m}( \bar{n}_{m}+1) \mathcal{\hat{D}}[ \hat{b}]  \hat{
\rho}^{\prime}+\gamma _{m} \bar{n}_{m}\mathcal{\hat{D}}[\hat{b}^{\dag }]  \hat{
\rho}^{\prime}.  \label{Eq9}
\end{eqnarray}%
In terms of Eq.~(\ref{Eq9}), we can derive the equations of motion for all the second-order moments of this three-mode optomechanical system (see Appendix). By solving these equations, we can obtain the mean phonon number in the MR.

\section{Adaibatic cooling of the mechanical mode}\label{sec:adiabatic}
\begin{table*}
\caption{Four kinds of coupling protocols for cooling the MR via the STIRAP: the Gaussian-, $\mathrm{sin^{4}}$-, $()^{-1/2}$-, and Vitanov-shaped coupling protocols.}
\label{tab:table1}
\begin{ruledtabular}
\begin{tabular}{ccccc}
$J(t)$  & $G_{2}(t)$  &$\dot{\theta}(t)$ &  $\mathrm{Ref.}$  \tabularnewline
\hline
 $ge^{-(t-t_{f}+\xi)/T}$  & $ge^{-(t-t_{f}-\xi)/T}$  & $-2\xi\{T^{2}\cosh[4\xi(t-t_{f})/T^{2}]\}^{-1}$ & {~\cite{Vitanov1997}}  \tabularnewline
 $g\sin^{4}[\pi(t+\xi)/T]$  & $g\sin^{4}(\pi t/T)$  & $-\frac{4\pi}{T}\sin(\frac{\pi\xi}{T})\sin^{3}(\frac{\pi t}{T})\sin^{3}[\frac{\pi(t+\xi)}{T}]\{\sin^{8}(\frac{\pi t}{T})+\sin^{8}[\frac{\pi(t+\xi)}{T}]\}^{-1}$ &{~\cite{Chen2010}}   \tabularnewline
 $g[1+e^{(t-t_{f})/T}]^{-1/2}$  & $g[1+e^{-(t-t_{f})/T}]^{-1/2}$  & $-\{4T\cosh[(t-t_{f})/2T]\}^{-1}$  & {~\cite{Laine1996}}  \tabularnewline
 $g\cos[\pi e^{t/T}(2e^{t/T}+2e^{10})^{-1}]$  & $g\sin[\pi e^{t/T}(2e^{t/T}+2e^{10})^{-1}]$  & $-\pi e^{t/T}e^{10}[2T(e^{t/T}+e^{10})^{2}]^{-1}$ & {~\cite{Vasilev2009}} \tabularnewline
\end{tabular}
\end{ruledtabular}
\end{table*}
In this section, we study how to cool a MR with the STIRAP method based on the approximate Hamiltonian~(\ref{eqA8}). In particular, we analyze the cooling process by mapping the three-mode system to a three-level system. The cooling process can be understood as an excitation transfer from mechanical mode $b$ to cavity mode $a_{1}$, which is equivalent to the population transfer in the three-level system with the STIRAP method~\cite{Gaubatz1990,Vitanov2017}. In order to implement the STIRAP process in our system, we first consider the case where the coupling between cavity mode $a_{1}$ and mechanical mode $b$ is turned off (i.e., $g_{1}=0$). Below, we assume the displacement amplitude $\alpha_{2}=\alpha_{2}^{*}$ for simplicity. In a rotating frame defined by the unitary operator $\exp [-i\omega _{m}t(\hat{a}_{1}^{\dag }\hat{a}_{1}+\hat{a}_{2}^{\dag }\hat{a}_{2}+\hat{b}^{\dag }\hat{b})]$ and within the rotating-wave approximation (RWA), the approximate Hamiltonian (\ref{eqA8}) is reduced to
\begin{equation}
\hat{H}_{\textrm{app}}'(t)=(\hat{a}_{1}^{\dag},\hat{a}_{2}^{\dag},\hat{b}^{\dag})
 \mathbf{M}(t)(\hat{a}_{1},\hat{a}_{2},\hat{b})^{\textrm{T}},  \label{eq:5}
\end{equation}
where the superscript ``$\textrm{T}$'' denotes the transpose of the matrix and we introduce the coupling matrix
\begin{equation}
\mathbf{M}(t)=\left(
\begin{array}{ccc}
0 & J\left( t\right)  & 0 \\
J\left( t\right)  & \delta  & G_{2}\left( t\right)  \\
0 & G_{2}\left( t\right)  & 0%
\end{array}\label{eq:6}
\right).
\end{equation}
In Eqs.~(\ref{eq:5}) and (\ref{eq:6}), we have considered the quasi-two-photon-resonance condition $\Delta _{1}=\omega _{1}-\omega _{L}=\omega _{m}$ and introduced the quasi-single-photon detuning $\delta =\Delta _{2}-\omega _{m}$. To clarify the physical mechanism for adiabatic cooling, we analyze the instantaneous eigensystems of the matrix $\mathbf{M}(t)$. To this end, we introduce the basis states $\vert a_{1}\rangle =(1,0,0) ^{\textrm{T}}$, $\vert a_{2}\rangle =(0,1,0) ^{\textrm{T}}$, and $\vert b\rangle =(0,0,1) ^{\textrm{T}}$; then we can understand the matrix $\mathbf{M}(t)$ as a time-dependent Hamiltonian of a three-level system with basis states $\vert a_{1}\rangle$, $\vert a_{2}\rangle$, and $\vert b\rangle$. The instantaneous eigenstates of the matrix $\mathbf{M}(t)$ can be obtained as
\begin{subequations}
\begin{eqnarray}
\!\vert \lambda _{0}(t)\rangle\!&=&\!\cos \theta\vert
a_{1}\rangle -\sin \theta \vert b\rangle , \\
\!\vert \lambda _{+}(t)\rangle\!&=&\!\sin \theta \sin \varphi \vert
a_{1}\rangle +\cos \varphi \vert a_{2}\rangle +\cos \theta
\sin \varphi \vert b\rangle , \   \notag \\
&& \\
\!\vert \lambda _{-}(t)\rangle\!&=&\!\sin \theta \cos \varphi \vert
a_{1}\rangle -\sin \varphi \vert a_{2}\rangle +\cos \theta
\cos \varphi \vert b\rangle ,\   \notag \\
\end{eqnarray}%
\end{subequations}
with the corresponding eigenvalues $E_{0}=0$, $E_{+}=g_{0}(t)\cot \varphi(t) $, and $E_{-}=-g_{0}(t)\tan
\varphi(t) $. Here, $g_{0}(t)=\sqrt{J^{2}( t) +G_{2}^{2}(t)}$ and the two mixing angles are defined by $\tan \theta(t) \equiv J(t)/G_{2}(t)$ and $\tan \varphi(t) \equiv g_{0}(t)/[\sqrt{\delta^{2}/4 +g_{0}^{2}(t)}+\delta /2] $. According to the mapping relation between the three-mode system and the three-level system, we investigate the excitation transfer in the three-mode system based on the physical mechanism for population transfer in the three-level system.

The population transfer from state $\vert b\rangle$ to state $\vert a_{1}\rangle$ can be realized by using the STIRAP protocols, i.e., the so-called counterintuitive modulation of the transition strengths~\cite{Gaubatz1990,Vitanov2017}. Here, the counterintuitive couplings satisfy the characteristic that the coupling strength $J(t)$ precedes $G_{2}(t)$. This can be described in an exact mathematical form as
\begin{equation}
\lim_{t\rightarrow 0}\frac{J(t) }{G_{2}(t) }=\infty ,%
\hspace{0.5cm}\lim_{t\rightarrow \infty }\frac{J(t) }{%
G_{2}(t) }=0\label{eq13}.
\end{equation}%
According to the definition of the mixing angle $\theta(t)$, Eq.~(\ref{eq13}) can also be expressed as
\begin{equation}
\theta (0) =\frac{\pi }{2},\hspace{0.5cm}\theta
(\infty )=0,
\label{eq8}
\end{equation}%
which means that at the initial time $t=0$, the adiabatic state is $\vert \lambda _{0}(0)\rangle =\vert b \rangle $, while at the ending time $t\rightarrow\infty$, the adiabatic state becomes $\vert \lambda _{0}(\infty)\rangle =\vert a_{1}\rangle $. If the population transfer process is adiabatic, the system will adiabatically follow the state $\vert \lambda _{0}\rangle$ all the time and, eventually, the population will be completely transferred from states  $\vert b \rangle$  to $\vert a_{1}\rangle $. It is worth noting that the theoretical analysis of the STIRAP process is based on the RWA. In fact, the counter-rotating terms will induce transitions involving states with different excitations. Particularly, the counter-rotating terms will increase excitations in this system, which leads to the heating of the system. Therefore, we need to choose proper parameter such that the influence of the counter-rotating terms can be neglected.

Corresponding to the three-mode system, we consider the case where the thermal phonon number in mechanical mode $b$ is $\bar{n}_{m}$ and the photon number in cavity mode $a_{1}$ is zero. After a perfect STIRAP process, these phonon excitations will be transferred to cavity mode $a_{1}$, and mechanical mode $b$ will be converted into a vacuum state. In particular, these excitations in cavity mode $a_{1}$ will be further extracted into its vacuum bath via photon loss. In this way, the thermal phonons in mechanical mode $b$ are extracted and then the mechanical mode is effectively cooled. Note that to confirm the adiabatic evolution, the adiabatic condition~\cite{Bergmann1998,Fleischhauer1996,Giannelli2014}
\begin{equation}
R(t)\equiv\frac{\vert \dot{\theta}(t)\vert }{\vert \delta/2\pm
\sqrt{\delta^{2}/4+g_{0}^{2}(t)}\vert}\ll1 \label{eq:7}
\end{equation}
should be satisfied.

The above discussions on the adiabatic cooling are based on an ideal case of the STIRAP. In a realistic case, the effect of the counter-rotating terms and system dissipations should be considered. The evolution of the mean phonon number $\langle\hat{b}^{\dagger}\hat{b}\rangle$, which is an indicator of the cooling, can be studied via the covariance matrix method. Based on the quantum master equation~(\ref{Eq9}), the equations of motion for all the second-order moments can be obtained, i.e., $\langle \hat{o}_{m}\hat{o}_{n}\rangle$ with $\hat{o}_{m}, \hat{o}_{n}$ $\in$ \{$\hat{a}_{i=1,2}$, $\hat{a}_{i}^{\dag }$, $\hat{b}$, and $\hat{b}^{\dag }$\}. Mathematically, these equations take the form (see the Appendix for equations of motion for all the second-order moments)
\begin{equation}
\frac{d\left\langle \hat{o}_{m}\hat{o}_{n}\right\rangle}{dt} =\mathrm{Tr}[\dot{\hat{\rho}}^{\prime}\hat{o}_{m}\hat{o}_{n}] =\sum_{i,j}\epsilon _{ij}\left\langle \hat{o}_{i}\hat{o}_{j}\right\rangle,
\label{eq10}
\end{equation}
where $\epsilon _{ij}$ is the corresponding coefficient. The initial conditions of these second-order moments can be determined based on the initial state of the system. In the following, we will consider an initial state of the system where only the mechanical mode $b$ is occupied, i.e., $\langle\hat{b}^{\dag}\hat{b}(0)\rangle$ is nonzero. In particular, the initial thermal occupations of the two cavity modes at room temperature are assumed to be vanishingly small. In this case, we assume that the initial mean phonon number of mode $b$ is $\langle\hat{b}^{\dag}\hat{b}(0)\rangle=10^{4}$, and that all other second-order moments are zero. With these initial conditions, the mean values of all the time-dependent second-order moments can be determined by solving Eq.~(\ref{eq10}). Based on the transient solution of Eq.~(\ref{eq10}), the time-dependent mean photon numbers $\langle\hat{a}^{\dag}_{i}\hat{a}_{i}(t)\rangle$ (for $i=1,2$) in the two cavity modes and the mean phonon number $\langle\hat{b}^{\dag}\hat{b}(t)\rangle$ in the mechanical mode can be obtained.

Note that the coefficients in these equations of motion for all the second-order moments depend on the photon-hopping strength $J(t)$ and the linearized optomechanical-coupling strengths $G_{1}(t)$ and $G_{2}(t)$. Therefore, the dynamics of the system can be controlled by choosing proper coupling strengths, which are determined by the pulsed driving fields.
\begin{figure*}[tbp]
\center \includegraphics[width=1.0\textwidth]{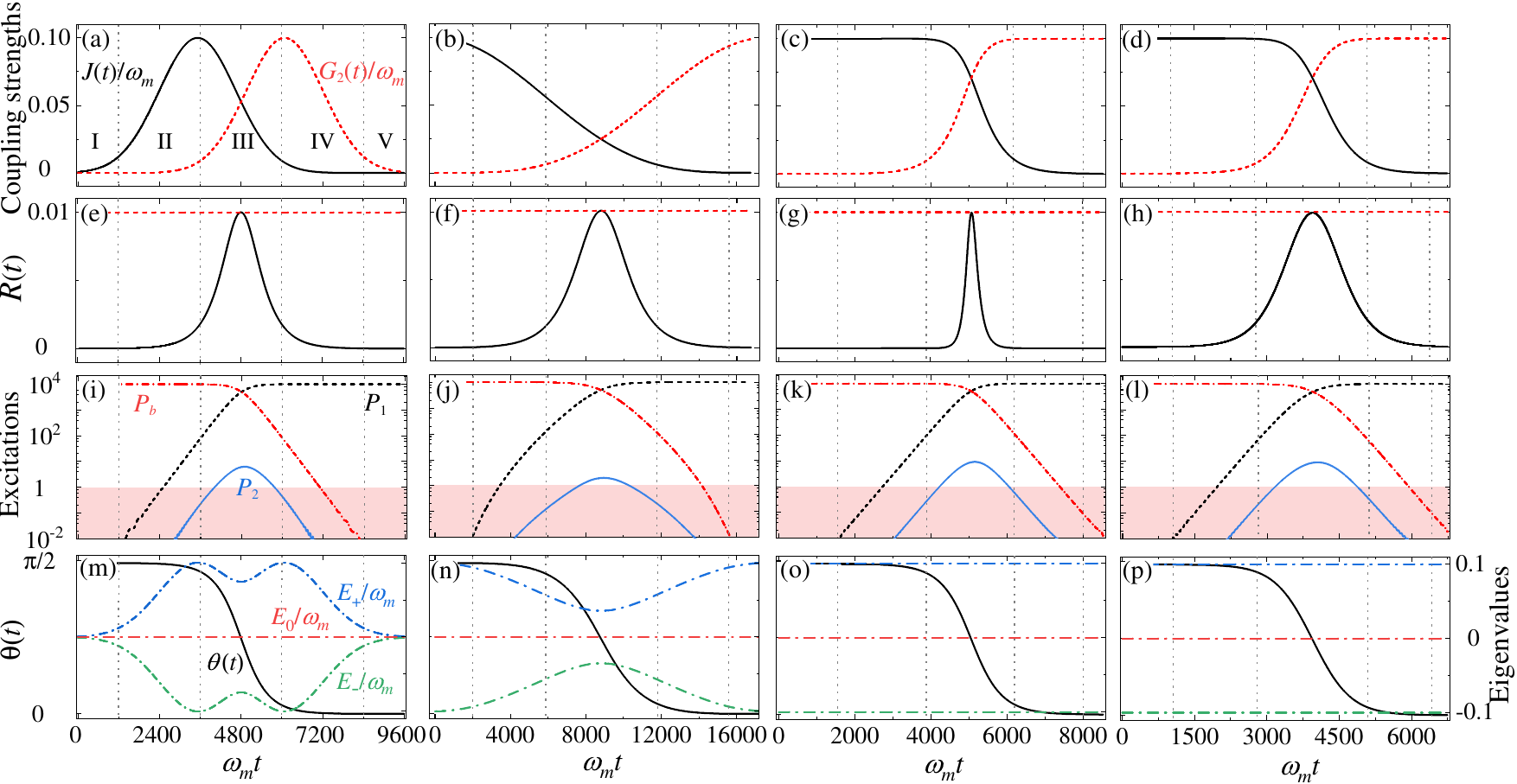}
\caption{(Color online) (a)-(d) The dimensionless coupling strengths $G_{2}(t)/\omega_{m}$ and $J(t)/\omega_{m}$ as functions of the scaled evolution time $\omega_{m}t$ in the adiabatic scheme. (e)-(h) Verification of the adiabatic condition given in Eq.~(\ref{eq:7}). (i)-(l) The mean photon numbers $P_{1}$ and $P_{2}$, and the mean phonon number $P_{b}$ vs the scaled evolution time $\omega_{m}t$ in the adiabatic scheme. (m)-(p) The mixing angle $\theta(t)$ and eigenvalues $E_{i=0,\pm}/\omega_{m}$ vs the scaled time $\omega_{m}t$. Here, the solid and dotted curves correspond to the left and right axes, respectively. Dotted vertical lines separate the five steps (labeled I-V) of the STIRAP scheme discussed in Sec.~\ref{sec:comparison}, and the five regions are distinguished by the ratio $J(t)/G_{2}(t)$. Panels in the first, second, third, and fourth columns correspond to the Gaussian-, $\mathrm{sin^{4}}$-, $()^{-1/2}$-, and Vitanov-shaped coupling cases, respectively. Other parameters used are $\delta=0$ and $g=0.1\omega_{m}$.}
\label{Fig2}
\end{figure*}

Below, we will consider four STIRAP protocols: the Gaussian~\cite{Vitanov1997}-, $\mathrm{sin^{4}}$~\cite{Chen2010}-, $()^{-1/2}$~\cite{Laine1996}-, and Vitanov~\cite{Vasilev2009}-shaped couplings, which are listed in Table~\ref{tab:table1}. For all these protocols, $g$ is the peak coupling strength, $T$ is the coupling pulse width, $\xi$ is the delay between the couplings $J(t)$ and $G_{2}(t)$, and $t_{f}$ is the time shift. In order to ensure that the STIRAP protocols approximately satisfy the condition in Eq.~(\ref{eq13}) at both the beginning and the ending of the protocols, we take $\xi=0.8T$ and $\xi=0.5T$ for the Gaussian- and $\mathrm{sin^{4}}$-shaped coupling protocols, respectively. We also take $t_{f}=3T$ and $t_{f}=20T$ for the Gaussian- and $()^{-1/2}$-shaped coupling protocols, respectively. To verify that the condition in Eq.~(\ref{eq13}) is well satisfied under our parameter conditions, in Figs.~\ref{Fig2}(a)--\ref{Fig2}(d) we show the dimensionless coupling strengths $J(t)/\omega_{m}$ and $G_{2}(t)/\omega_{m}$ versus the scaled evolution time $\omega_{m}t$. Here, we can see that $J(t)$ (solid black curves) precedes $G_{2}(t)$ (dashed red curves), which is a characteristic of counterintuitive modulation. More specifically, we have $J(t_{s})/G_{2}(t_{s})\approx 1.47\times10^{4}, 6.44\times10^{5}, 2.2\times10^{4}$, and $1.4\times10^{4}$ [$J(t_{e})/G_{2}(t_{e})\approx 6.50\times10^{-4}, 4\times10^{-5}, 1.04\times10^{-3}$, and $1.32\times10^{-3}$] for different STIRAP protocols, where $t_{s}$ and $t_{e}$ represent the moment at the beginning and the ending of the STIRAP protocols, respectively. The result shows that the condition in Eq.~(\ref{eq13}) is satisfied well under our parameters.

Additionally, all these protocols need to satisfy the adiabatic condition which is related to the parameters $\delta$, $g$, and $T$. For fixed  parameters $\delta$ and $g$, the larger $T$ is, the better the adiabatic criteria in Eq.~(\ref{eq:7}) are satisfied. In this paper, we consider that the maximum value of $R(t)$ defined in Eq.~(\ref{eq:7}) is less than $0.01$ and satisfies the adiabatic condition. Thus, for the given parameters $\delta=0$ and $g=0.1\omega_{m}$, the values of $T$ are $1600\omega_{m}^{-1}$, $35200\omega_{m}^{-1}$,  $253\omega_{m}^{-1}$, and $395\omega_{m}^{-1}$ for the Gaussian-, $\mathrm{sin^{4}}$-, $()^{-1/2}$-, and Vitanov-shaped coupling protocols, respectively. At the same time, we plot the dependence of the factor $R(t)$ on the scaled evolution time $\omega_{m}t$ in Figs.~\ref{Fig2}(e)--\ref{Fig2}(h). The maximum value of $R(t)$ (solid black curves) is less than $0.01$ (dashed red curves), which means that the adiabatic condition is well satisfied at this time.

To demonstrate the implementation of the STIRAP process in our three-mode optomechanical system, we show in Figs.~\ref{Fig2}(i)--\ref{Fig2}(l) the process of adiabatic excitation transfer without considering the system dissipations. To characterize the cooling advantage, we choose the moment when the mean phonon number first reaches the minimum value less than 1 as the reference time of the cooling performance, and the corresponding times are $8473\omega_{m}^{-1}$, $16750\omega_{m}^{-1}$, $8531\omega_{m}^{-1}$, and $6746\omega_{m}^{-1}$ for the Gaussian-, $\mathrm{sin^{4}}$-, $()^{-1/2}$-, and Vitanov-shaped coupling protocols, respectively. Here, we can see that a perfect excitation transfer from mechanical mode $b$ (dash-dotted red curves) to cavity mode $a_{1}$ (dashed black curves) is realized through the STIRAP process. In particular, at the ending of the pulsed driving field, the mean phonon number $P_{b}=\textrm{Tr}[\hat{b}^{\dagger}\hat{b}\hat{\rho}'(t)]$ in the mechanical mode is less than $1$ ($0.0033$, $0.0025$, $0.013$, and $0.0185$ for the Gaussian-, $\mathrm{sin^{4}}$-, $()^{-1/2}$-, and Vitanov-shaped coupling protocols, respectively), which means that the ground-state cooling of the MR can be realized with the STIRAP method. The small value of the excitation number $P_{2}=\textrm{Tr}[\hat{a}_{2}^{\dagger}\hat{a}_{2}\hat{\rho}'(t)]$ in intermediate mode $a_{2}$ indicates that the cooling process is robust against the dissipation from the intermediate mode by following the dark state $\vert \lambda _{0}(t)\rangle$ adiabatically. We point out that compared with the chirped pulse scheme~\cite{Liao2011,Chen2015}, it is easier to choose parameters for the STIRAP scheme because these parameters can be confirmed by the adiabatic condition.

\section{Comparison of the four coupling protocols\label{sec:comparison}}

We have discussed ground-state cooling of the MR using the STIRAP method in Sec.~\ref{sec:adiabatic} and found that the dynamic cooling processes are different for four kinds of coupling protocols. In this section, we will clarify the differences among these four coupling protocols and draw some conclusions concerning the features of the transient cooling. For these four kinds of coupling protocols, the mechanism of the STIRAP can be understood by dividing the interaction process into five steps~\cite{Fleischhauer2001}, delineated by the dashed vertical lines in Fig.~\ref{Fig2} and distinguished by the ratio $J(t)/G_{2}(t)$.

\emph{Step I}. In this step, the coupling strength $G_{2}(t)$ is absent in these four cases. The coupling strength $J(t)$ increases from zero for the Gaussian-shaped protocol. However, for the other three cases, the coupling strength $J(t)$ starts from a finite value. Concretely, $J(t)$ decreases for the $\mathrm{sin^{4}}$-shaped protocol and remains unchanged for the other two cases. In step I, the system will remaim in the adiabatic eigenstate $|\lambda_{0}(t)\rangle$ [$J(t)/G_{2}(t)\rightarrow\infty$ corresponding to $\theta(t)\rightarrow\pi/2$]. The nonzero eigenvalues of the system are $E_{\pm}=\pm \sqrt{J^{2}( t) +G_{2}^{2}(t)}$ at $\delta=0$. Therefore, $E_{\pm}(t)$ and $J(t)$ have the same evolution trend. For the Gaussian ($\mathrm{sin^{4}}$)-shaped protocol, the adiabatic energy separation (AES) increases (decreases) due to the increase (decrease) of the coupling strength $J(t)$. For the $()^{-1/2}$- and Vitanov-shaped protocols, the AES remains unchanged.

\emph{Step II}. During this step, the coupling strength $J(t)$ maintains the same evolution trend as that in step I, and the coupling strength $G_{2}(t)$ begins to increase. In this step, the average photon number $P_{1}$ is slowly added and the AES is still dominated by the coupling strength $J(t)$ because $J(t)\gg G_{2}(t)$. For the Gaussian ($\mathrm{sin^{4}}$)-shaped protocol, the AES continues to raise (reduce) due to the increase (decrease) of the coupling strength $J(t)$. For the $()^{-1/2}$- and Vitanov-shaped protocols, the AES is nearly unchanged due to the negligible change of the coupling strength $J(t)$.

\emph{Step III}. In this step, the coupling strength $G_{2}(t)$ increases but $J(t)$ decreases. Consequently, the mixing angle $\theta(t)$ decreases from $\pi/2$ to 0, and the state vector $|\psi(t)\rangle=|\lambda_{0}(t)\rangle$ evolves adiabatically from $-|b\rangle$ to $|a_{1}\rangle$. In this step, the AES is determined by both coupling strengths $J(t)$ and $G_{2}(t)$. For the Gaussian and $\mathrm{sin^{4}}$-shaped protocols, the AES experiences an anti-crossing. For the $()^{-1/2}$- and Vitanov-shaped protocols, the AES remains unchanged.

\emph{Step IV}. In the fourth step, the coupling strength $G_{2}(t)$ plays a dominant role in the AES because the coupling strength $G_{2}(t)$ is much greater than $J(t)$. For the Gaussian ($\mathrm{sin^{4}}$)-shaped protocol, the AES decreases (increases) due to the decrease (increase) of the coupling strength $G_{2}(t)$. For the $()^{-1/2}$- and Vitanov-shaped protocols, the AES remains unchanged. The mixing angle $\theta(t)\rightarrow0$ [$G_{2}(t)/J(t)\gg 0$] and the state vector $|\psi(t)\rangle$ is almost in target state $|a_{1}\rangle$, which means that the excitations are almost deposited in mode $a_{1}$.

\emph{Step V}. In the last step, only the coupling strength $G_{2}(t)$ exists, and the state vector $|\psi(t)\rangle$ is right on the target state $|a_{1}\rangle$. For the ``Gaussian'' (``$\mathrm{sin^{4}}$'')-shape protocol, the AES gradually reduces (raises) to 0 (0.2$\omega_{m}$) as the coupling strength $G_{2}(t)$ reduces (raises) to 0 (0.1$\omega_{m}$). For the ``$()^{-1/2}$''- and ``Vitanov''-shape protocols, the AES remains unchanged.

In general, the adiabatic evolution of these four pulses in the STIRAP process requires the AES, which induces the excitation transfer~\cite{Sarma2020NJP,Stefano2015}. However, the evolution of the AES is different for these four cases. For the Gaussian- and $\mathrm{sin^{4}}$-shaped protocols, the AES experiences an anti-crossing, while for the $()^{-1/2}$- and Vitanov-shaped couplings, the AES remains unchanged. In addition, it can be seen from Figs.~\ref{Fig2}(a)--\ref{Fig2}(d) that the Gaussian-shaped protocol has vanishing coupling strengths at the initial and final moments, while for the $\mathrm{sin^{4}}$-, $()^{-1/2}$-, and Vitanov-shaped couplings, the coupling strengths are nonzero. We also find that the excitation transfer for these four protocols starts from the beginning of the coupling strength $G_{2}(t)$ and finishes with the completion of the coupling strength $J(t)$. Moreover, the mixing angle $\theta(t)$ has a similar evolution trend.

\section{Accelerated cooling of the mechanical mode via shortcuts to adiabaticity\label{sec:STA}}

\begin{figure*}[tbp]
\center \includegraphics[width=1.0\textwidth]{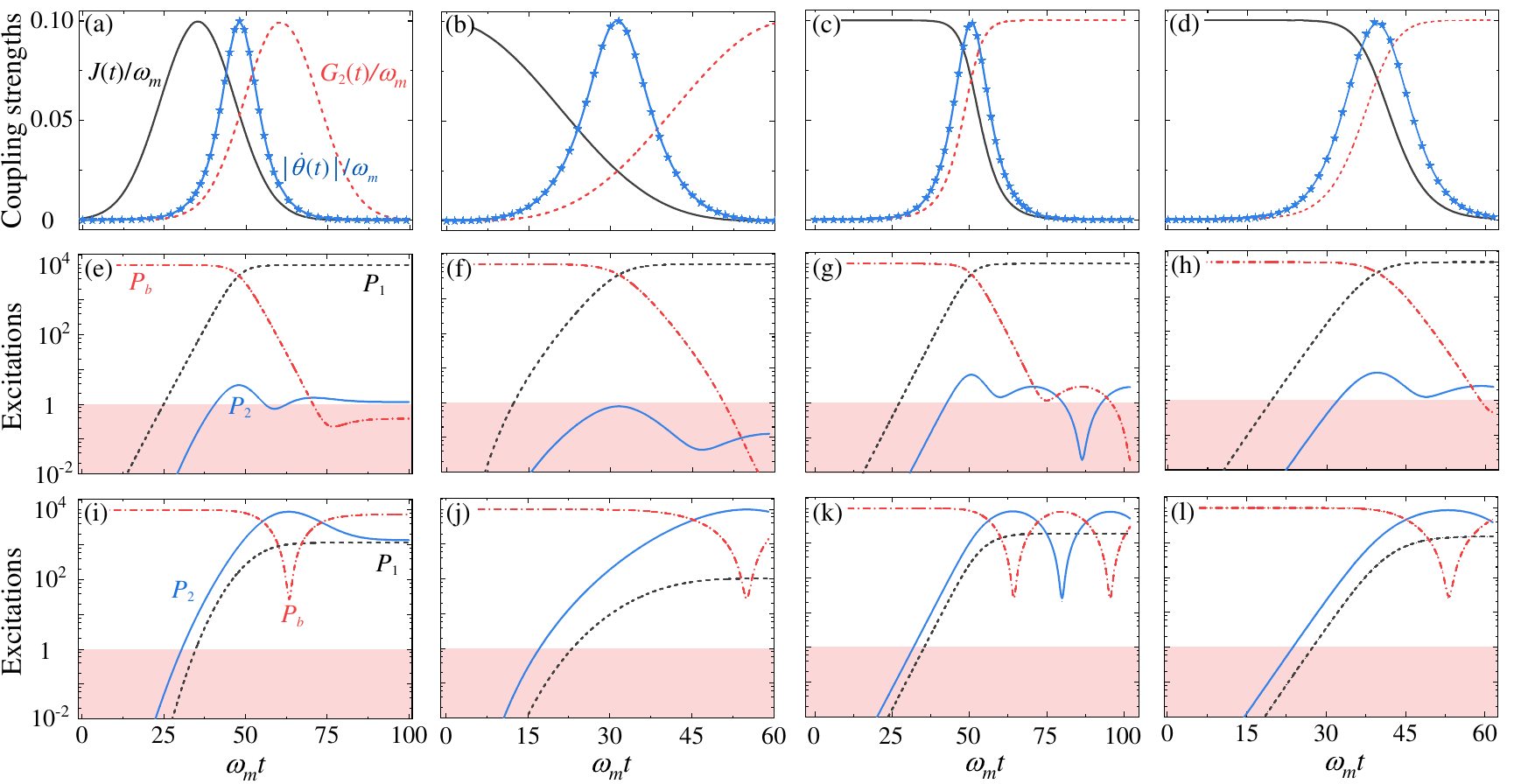}
\caption{(Color online) (a)-(d) The dimensionless coupling strengths $G_{2}(t)/\omega_{m}$ and $J(t)/\omega_{m}$, and the dimensionless counter-adiabatic interaction term $|\dot{\theta}(t)|/\omega_{m}$ as functions of the scaled evolution time $\omega_{m}t$ in the STA scheme. (e)-(h) The time evolution of the mean photon numbers $P_{1}$ and $P_{2}$, and the mean phonon number $P_{b}$ in the STA scheme. (i)-(l) Dynamics of the mean photon numbers $P_{1}$ and $P_{2}$, and the mean phonon number $P_{b}$ in the absence of the counter-adiabatic interaction term. Panels in the first, second, third, and fourth columns correspond to the Gaussian-, $\mathrm{sin^{4}}$-, $()^{-1/2}$-, and Vitanov-shaped coupling cases, respectively. Other parameters used are $\delta=0$ and $g=0.1\omega_{m}$.}
\label{Fig3}
\end{figure*}
In this section, we study accelerated ground-state cooling of the MR with the STA method. Motivated by the counteradiabatic driving scheme in the three-level atomic system, we introduce the following counteradiabatic interaction term ~\cite{Berry2009,Chen2010}
\begin{equation}
\mathbf{M}_{cd}(t)=i\sum_{n=0,\pm }\partial _{t}\left\vert \lambda _{n}\left(
t\right) \right\rangle \left\langle \lambda _{n}\left( t\right) \right\vert ,
\end{equation}%
which can implement counteradiabatic transitions in a system described by a Hamiltonian with the matrix form $\mathbf{M}(t)$ in Eq.~(\ref{eq:6}). Accordingly, for our three-mode optomechanical system, the introduced interaction for implementing the counteradiabatic process takes the form
\begin{equation}
\hat{H}_{cd}(t)=i(\hat{a}_{1}^{\dag},\hat{a}_{2}^{\dag},\hat{b}^{\dag})
 \mathbf{M}_{cd}(t)(\hat{a}_{1},\hat{a}_{2},\hat{b})^{\textrm{T}},
\end{equation}
with
\begin{equation}
\mathbf{M}_{cd}(t)=
\left(
\begin{array}{rrr}
0 & \dot{\varphi}\sin \theta & \dot{\theta} \\
-\dot{\varphi}\sin \theta & 0 & -\dot{\varphi}\cos \theta \\
-\dot{\theta} & \dot{\varphi}\cos \theta & 0%
\end{array}%
\right),
\end{equation}
where $\dot{\theta}=[\dot{J}(t)G_{2}(t)-J(t)\dot{G}_{2}(t)]/[J^{2}\left(
t\right) +G_{2}^{2}(t)]$ and $\dot{\varphi}=\{[\dot{J}(t)J(t)+\dot{G}_{2}(t)G_{2}(t)]\delta\}/\{[\delta ^{2}+4g_{0}^{2}(t)]g_{0}(t)\}$. In principle, we would need three new interactions to implement this Hamiltonian. By working in the adiabatic basis we see that $d\langle \lambda _{0}(t) \vert \psi ^{I}( t) \rangle /dt$ is independent of $\dot{\varphi}$ for arbitrary $|\psi ^{I}(t)\rangle$. Therefore, $d\langle \lambda _{0}(t) \vert \psi ^{I}( t) \rangle /dt$ is immune to both the $a_{1}$-$a_{2}$ and $a_{2}$-$b$ auxiliary interactions, which thus are unnecessary for a full passage from mechanical mode $b$ to cavity mode $a_{1}$. As a result, matrix $\mathbf{M}_{cd}$ may be simplified as~\cite{Chen2010}
\begin{equation}
\mathbf{M}'_{cd}(t)=%
\begin{pmatrix}
0 & 0 & \dot{\theta} \\
0 & 0 & 0 \\
-\dot{\theta} & 0 & 0%
\end{pmatrix},
\end{equation}%
and the corresponding Hamiltonian is denoted as $\hat{H}'_{cd}(t)=i(\hat{a}_{1}^{\dag},\hat{a}_{2}^{\dag},\hat{b}^{\dag})\mathbf{M}'_{cd}(t)(\hat{a}_{1},\hat{a}_{2},\hat{b})^{\textrm{T}}$. Based on Eq.~(\ref{eq:5}) and $\hat{H}'_{cd}(t)$, the total Hamiltonian can be written as
\begin{eqnarray}
\hat{H}\left( t\right) &\mathcal{=}&\hat{H}_{\textrm{app}}'(t) +%
\hat{H}'_{cd}(t)  \notag \\
&=& (\hat{a}_{1}^{\dag},\hat{a}_{2}^{\dag},\hat{b}^{\dag})
\mathbf{N}(t)(\hat{a}_{1},\hat{a}_{2},\hat{b})^{\textrm{T}},  \label{Eq13}
\end{eqnarray}%
with
\begin{equation}
\mathbf{N}(t)=\left(
\begin{array}{rrr}
0 & J\left( t\right) & i\dot{\theta} \\
J\left( t\right) & \delta & G_{2}\left( t\right) \\
-i\dot{\theta} & G_{2}\left( t\right) & 0%
\end{array}%
\right).
\end{equation}
The counteradiabatic interaction term is a direct coupling between cavity mode $a_{1}$ and mechanical mode $b$, which means that the model we proposed in Fig.~\ref{Fig1} can be used to accelerate the ground-state cooling of a MR.

\begin{figure*}[tbp]
\center \includegraphics[width=1.0\textwidth]{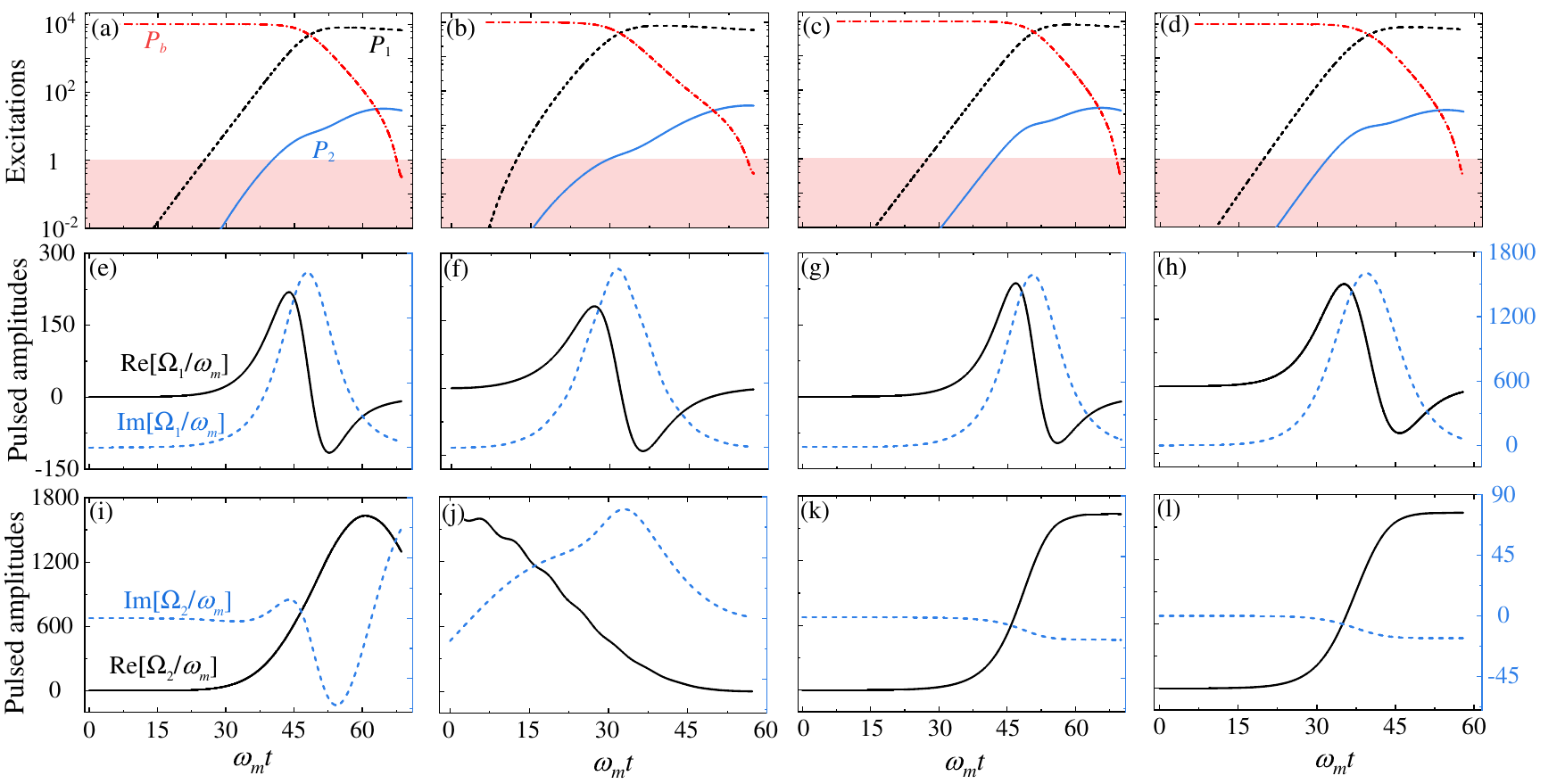}
\caption{(Color online) (a)-(d) The mean photon numbers $P_{1}$ and $P_{2}$, and the mean phonon number $P_{b}$ vs the scaled evolution time $\omega_{m}t$ in the dissipative case. The original driving amplitudes (e)-(h) $\Omega_{1}(t)$ and (i)-(l) $\Omega_{2}(t)$ vs the scaled time $\omega_{m}t$ in the dissipative case. Here, the solid and dashed curves correspond to the left and right axes, respectively. Panels in the first, second, third, and fourth columns correspond to the Gaussian-, $\mathrm{sin^{4}}$-, $()^{-1/2}$-, and Vitanov-shaped coupling cases, respectively. Other parameters used are $\delta=0$, $g/\omega_{m}=0.1$, $\kappa_{1}/\omega_{m}=2\times10^{-2}$, $\kappa_{2}/\omega_{m}=2\times10^{-2}$, $\gamma_{m}/\omega_{m}=3\times10^{-6}$, $g_{1}/\omega_{m}=6\times10^{-5}$, and $g_{2}/\omega_{m}=6\times10^{-5}$.}
\label{Fig4}
\end{figure*}

In principle, $\hat{H}'_{cd}(t)$ drives the dynamics in any short times along the adiabatic path of $\hat{H}_{\textrm{app}}'(t)$, but there are practical limitations such as available pulse driving power. Moreover, a comparison with $\hat{H}_{\textrm{app}}'(t)$ dynamics is only fair if $|\dot{\theta}(t)|$ is smaller or approximately equal to the peak coupling strength $g$, which means that the counter-adiabatic interaction term must satisfy the inequality $|\dot{\theta}(t)|\leq g$~\cite{Chen2010} [for the four kinds of STIRAP protocols considered in this work, the corresponding $\dot{\theta}(t)$ is given in the third column of Table~\ref{tab:table1}]. This inequality implies that we can speed up the STIRAP process with a minimal time $T$ under the given parameters $\delta=0$ and $g=0.1\omega_{m}$. Through numerical verification, we choose $16\omega_{m}^{-1}$, $126\omega_{m}^{-1}$, $2.53\omega_{m}^{-1}$, and $3.95\omega_{m}^{-1}$ for the Gaussian-, $\mathrm{sin^{4}}$-, $()^{-1/2}$-, and Vitanov-shaped coupling protocols, respectively. We show in Figs.~\ref{Fig3}(a)--\ref{Fig3}(d) the shapes of the dimensionless coupling strengths $J(t)/\omega_{m}$ and $G_{2}(t)/\omega_{m}$, and the dimensionless counteradiabatic interaction term $|\dot{\theta}(t)|/\omega_{m}$ as functions of the scaled evolution time $\omega_{m}t$. The value of $|\dot{\theta}(t)|/\omega_{m}$ (blue star curves) is always less than or equal to $0.1$ ($g/\omega_{m}=0.1$), which means that the values of the parameter $T$ used for different protocols satisfy the inequality $|\dot{\theta}(t)|\leq g$. In addition, at the end of the pulse, the value of $\dot{\theta}$ becomes zero, which indicates that the counter-adiabatic interaction term $\dot{\theta}$ only exists in the accelerated cooling process and has no effect on the subsequent excitation evolution.

To demonstrate that the STA scheme corrects for the nonadiabatic losses even when the adiabatic condition for the STIRAP method is not satisfied, we plot the evolution of the mean photon and phonon numbers versus the scaled evolution time $\omega_{m}t$ in Figs.~\ref{Fig3}(e)--\ref{Fig3}(h). We can see that the phonon excitations (dash-dotted red curves) in mechanical mode $b$ are transferred to the photon excitations (dashed black curves) in cavity mode $a_{1}$ in a much shorter period. In particular, the mean phonon number $P_{b}$ in the mechanical mode at the end of the pulsed driving field ($77\omega_{m}^{-1}$, $59\omega_{m}^{-1}$, $102\omega_{m}^{-1}$, and $61.5\omega_{m}^{-1}$ for the Gaussian-, $\mathrm{sin^{4}}$-, $()^{-1/2}$-, and Vitanov-shaped coupling protocols, respectively) is less than $1$ ($0.23$, $0.0029$, $0.022$, and $0.46$ for the Gaussian-, $\mathrm{sin^{4}}$-, $()^{-1/2}$-, and Vitanov-shaped coupling protocols, respectively), which means that the fast ground-state cooling of the MR is realized with the STA method. Here, the cooling performance and efficiency are limited by the pulse driving power, i.e., the parameter $T$. The smaller the value of $T$, the faster and more efficient cooling of the MR will be realized. Compared with the results shown in Figs.~\ref{Fig2}(i)--\ref{Fig2}(l), the cooling of the MR is accelerated by about $110$, $283$,  $83.6$, and $109.7$ times for the Gaussian-, $\mathrm{sin^{4}}$-, $()^{-1/2}$-, and Vitanov-shaped coupling protocols, respectively.  To better illustrate the effect of the counter-adiabatic interaction term, we also show the evolution of the mean photon and phonon numbers in the absence of the counter-adiabatic interaction terms in Figs.~\ref{Fig3}(i)--\ref{Fig3}(l). The results indicate that the counter-adiabatic interaction term plays an important and necessary role in the fast ground-state cooling of the MR.

As mentioned before, the interactions with environments will inevitably lead to the dissipation of the system. In Figs.~\ref{Fig4}(a)--\ref{Fig4}(d), we plot the evolution of the mean photon and phonon numbers versus the scaled evolution time $\omega_{m}t$ in the dissipative case. Here, we can see that the mean phonon number $P_{b}$ in the mechanical mode decreases rapidly from its initial value ($\bar{n}_{m}=10^{4}$) to a relatively small number (about 0.324, 0.331, 0.389, and 0.323 for the Gaussian-, $\mathrm{sin^{4}}$-, $()^{-1/2}$-, and Vitanov-shaped coupling protocols, respectively) in a short time duration. These indicate that the ground-state cooling of the MR can be realized for four kinds of coupling protocols.

We note that in our parameter conditions, the system will enter the strong-coupling regime at some times (around the peak value couplings). It has been found in previous studies~\cite{Dobrindt2008PRL,He2017PRL} that for a single optomechanical cavity driven by a continuous wave, the cooling limit of the MR is $n_{f}=\bar{n}_{m}\gamma_{m}/\kappa_{2}$ in the strong-coupling regime. For our used parameters, the cooling limit corresponding to the continuous-wave driving and strong-coupling regime is $n_{f}=1.5$. In this sense, the cooling results ($0.324, 0.331, 0.389, 0.323 <1.5$) that we get can surpass the single-cavity cooling limit in the strong-coupling regime. However, the cooling limit obtained in previous studies corresponds to the case of a single optomechanical cavity driven by a continuous field. It has been found that the cooling performance for the square pulse driving case can surpass the theoretical cooling limit corresponding to the case of continuous-wave driving with the same amplitude~\cite{LinOE2018}. This indicates that the pulse driving could break the cooling limit of the continuous-wave driving in proper cases. In addition, we consider the cooling of a single mechanical resonator coupled to two optical cavities and the system forms a looped coupling. In this system, the quantum interference effect will affect the cooling performance of the mechanical resonator under certain conditions~\cite{LiuPRA2015,WangPRA2019}.

We also investigate the pulsed driving fields corresponding to the four cases of coupling protocols. We show in Figs.~\ref{Fig4}(e)--\ref{Fig4}(h) [Figs.~\ref{Fig4}(i)--\ref{Fig4}(l)] the original driving amplitude $\Omega_{1}(t)$ [$\Omega_{2}(t)$] versus the scaled evolution time $\omega_{m}t$. The original driving amplitudes $\Omega_{1}(t)$ and $\Omega_{2}(t)$ are obtained by solving Eqs.~(\ref{5})-(\ref{6}) under the couplings $G_{1}(t)$ and $G_{2}(t)$. It can be seen that all the pulsed driving fields have smooth shape and reasonable amplitudes, which confirm the experimental feasibility of our schemes.

Now, we compare the four coupling protocols in the case of STA. We can see from Figs.~\ref{Fig3}(a)--~\ref{Fig3}(d) that the counteradiabatic interaction terms $\dot{\theta}(t)$ among the four protocols are similar but not identical. For Gaussian- and $()^{-1/2}$-shaped protocols, the counteradiabatic interaction term is zero for a long time after the beginning and before the end of the protocol, which obviously leads to an increase of the cooling time, while for $\mathrm{sin^{4}}$- and Vitanov-shaped protocols, this period is very short, thus shortening the cooling time. This inspires us that in the STA case we are able to shorten this time to quicken the cooling. Moreover, it can be seen from Figs.~\ref{Fig4}(e)--~\ref{Fig4}(h) [Figs.~\ref{Fig4}(i)--~\ref{Fig4}(l)] that the original driving amplitudes $\Omega_{1}(t)$ [$\Omega_{2}(t)$] are different for the four protocols in the STA case when the system dissipations are included.

\begin{figure}[tbp]
\center \includegraphics[width=0.45\textwidth]{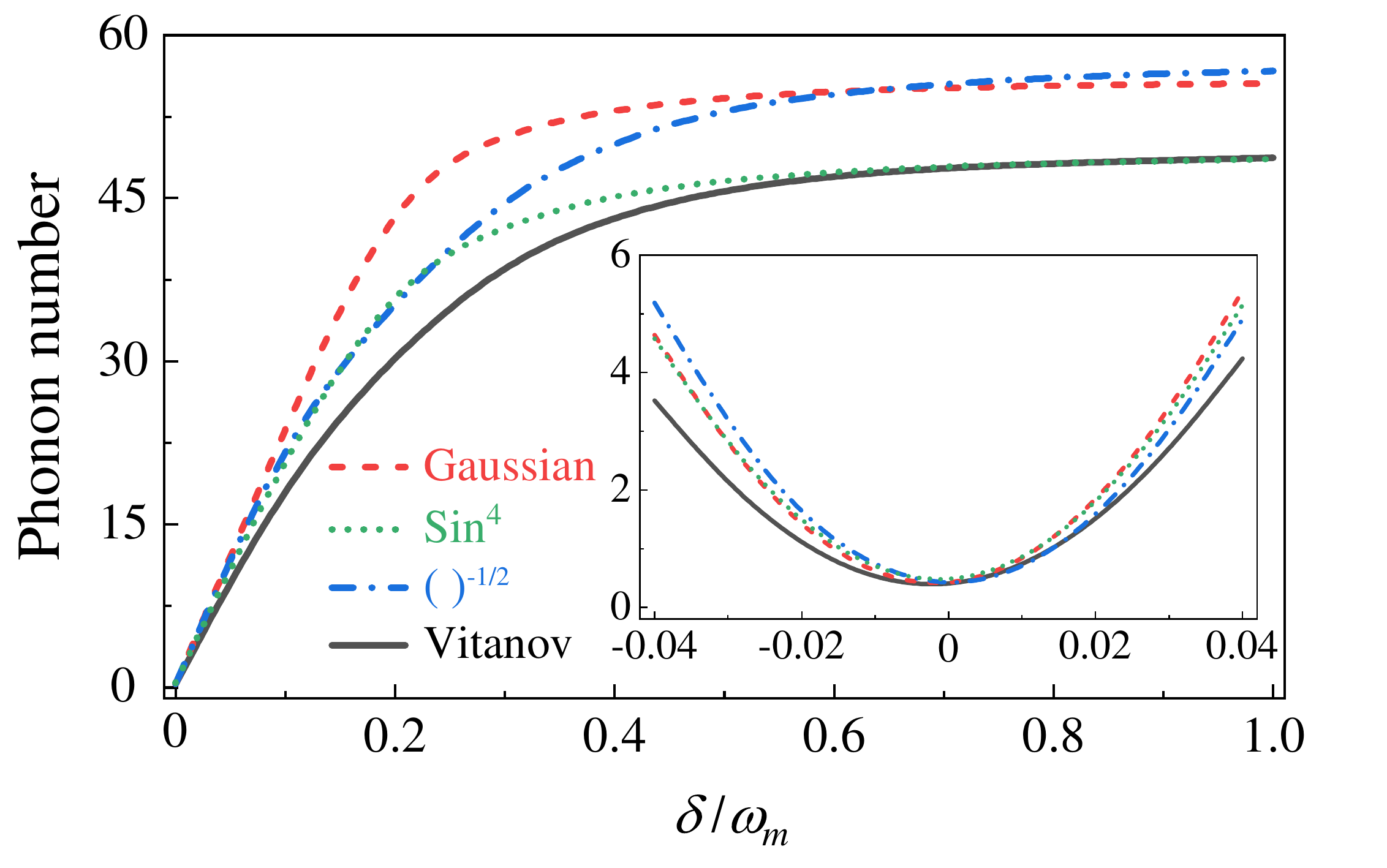}
\caption{(Color online) Plot of the mean phonon number (at the ending of the pulsed driving fields) as a function of the scaled quasi- single-photon detuning $\delta/\omega_{m}$. Inset: The mean phonon number vs $\delta$ around $0$. Other parameters used are $g/\omega_{m}=0.1$, $\kappa_{1}/\omega_{m}=2\times10^{-2}$, $\kappa_{2}/\omega_{m}=2\times10^{-2}$, and $\gamma_{m}/\omega_{m}=3\times10^{-6}$.}
\label{Fig5}
\end{figure}

In our previous simulations, we considered the quasi-single-photon-resonance case, $\delta=0$. Actually, the quasi-single-photon detuning is a tunable parameter. Therefore, it is an interesting question to find how the cooling depends on the quasi single-photon detuning $\delta$. In Fig.~\ref{Fig5}, we investigate the mean phonon number (at the end of the pulsed driving field) as a function of the scaled quasi-single-photon detuning $\delta/\omega_{m}$ to seek an optimal detuning. Here, we can see that the mean phonon number increases with the increase of the quasi-single-photon detuning $\delta$ for four kinds of protocols, which shows that the quasi-single-photon resonance is an optimal point. The inset shows the mean phonon number near $\delta=0$. Here, we can see that the four kinds of protocols are sensitive to the detuning.

\section{Discussions on the experimental implementation\label{sec:EXPERIMENT}}

In this section, we present some discussions of the experimental implementation of this scheme. In this work, there are two kinds of interactions in the system, namely, the time-dependent linearized optomechanical interaction between the cavity field and the MR, and the time-dependent photon-hopping interaction between the two cavity fields. It is worth mentioning that the time-dependent linearized optomechanical interaction $G_{i=1,2}(t)=g_{i}\alpha_{i}(t)$ can be adjusted by controlling the pulsed fields~\cite{Tian2015}. In addition, the photon hopping between the superconducting resonators in electromechanical systems via the Josephson junction coupling has been proposed theoretically~\cite{Devoret1997,Liao2016}, which indicates the feasibility of the experimental implementation of the time-dependent photon-hopping interaction. Note that the time-dependent photon-hopping interaction has been widely used in synthetic dimension in cavity systems, and the experimental implementation of this kind of interaction has been discussed in detail~\cite{Fang2012Photonic,Fang2012PRL}.

Below, we focus our discussions on the electromechanical systems because it is possible to realize the time-dependent photon-hopping interaction in this setup. The realistically experimental parameters in electromechanical systems to realize the ground-state cooling of the MR are as follows~\cite{Teufel2011Nature}: $\omega_{m}=2\pi\times10.56$ MHz, $\gamma_{m}=2\pi\times32$ Hz, $\kappa_{i=1,2}=2\pi\times200$ kHz, and $g_{i=1,2}=2\pi\times200.9\pm8.2$ Hz. Taking $\omega_{m}$ as the frequency scale, we have $\gamma_{m}/\omega_{m}\approx3.03\times10^{-6}$, $\kappa_{i=1,2}/\omega_{m}\approx1.894\times10^{-2}$, and $g_{i=1,2}/\omega_{m}\approx1.824$-$1.980\times10^{-5}$. The parameter conditions for implementation of our scheme discussed above
are $\kappa_{i=1,2}/\omega_{m}=2\times10^{-2}$, $\gamma_{m}/\omega_{m}=3\times10^{-6}$, and $g_{i=1,2}/\omega_{m}=6\times10^{-5}$, which are at the same order as the experimental parameters in Ref.~\cite{Teufel2011Nature}. Therefore, the system parameters used in this work should be within the reach of current experimental conditions. Under these parameter conditions, the shapes of the original driving amplitudes $\Omega_{1}(t)$ and $\Omega_{2}(t)$ are shown in Figs.~\ref{Fig4}(e)--\ref{Fig4}(h) and Figs.~\ref{Fig4}(i)--\ref{Fig4}(l), respectively. The smooth shape and moderate magnitudes of the pulsed field also confirm the experimental feasibility of our schemes.

\section{CONCLUSION\label{sec:conclusion}}

In conclusion, we have proposed a transient-state-cooling scheme based on the STA method to realize fast ground-state cooling of a MR. We have considered four kinds of coupling protocols and compared the cooling efficiency of the MR based on the STIRAP method and the STA method. We have verified that the ground-state cooling of the MR can be realized with the STA method, and the cooling velocities are increased by nearly two orders of magnitude. We have also shown that the original driving amplitudes of the pulsed fields have a smooth shape and moderate magnitudes, which confirm the experimental feasibility of our schemes. The STA method for fast ground-state cooling of the MR will perhaps inspire new ideas for accelerating other adiabatic evolution processes in optomechanical systems, and shed light on the development of fast optomechanical quantum operations.

\begin{acknowledgments}
J.-Q.L. thanks Professor Bing He for valuable discussions. J.-Q.L. is supported in part by the National Natural Science Foundation of China (Grants No.~12175061, No.~11774087, No.~11822501, and No.~11935006), Hunan Science and Technology Plan Project (Grant No.~2017XK2018), and the Science and Technology Innovation Program of Hunan Province (Grants No.~2020RC4047 and No.~2021RC4029). J.-F.H. is supported in part by the National Natural Science Foundation of China (Grant No. 12075083), Scientific Research Fund of Hunan Provincial Education Department (Grant No. 18A007), and Natural Science Foundation of Hunan Province, China (Grant No. 2020JJ5345).
\end{acknowledgments}

\appendix*
\begin{widetext}
\section{The equation of motion for all the second-order moments \label{appendix}}
In this Appendix, we present the equations of motion for all of the second-order moments, which are obtained based on Eq.~(\ref{Eq9}) as
\begin{eqnarray}
\frac{d}{dt}\langle \hat{a}_{1}^{\dag }\hat{a}_{1}\rangle  &=&-iJ(t)\langle \hat{a}_{1}^{\dag
}\hat{a}_{2}\rangle +iJ(t)\langle \hat{a}_{1}\hat{a}_{2}^{\dag }\rangle +iG_{1}(t)\langle
\hat{a}_{1}\hat{b}^{\dag }\rangle -iG_{1}^{\ast }(t)\langle \hat{a}_{1}^{\dag }\hat{b}\rangle
-\kappa_{1}\langle \hat{a}_{1}^{\dag }\hat{a}_{1}\rangle ,  \notag\\
\frac{d}{dt}\langle \hat{a}_{2}^{\dag }\hat{a}_{2}\rangle  &=&iJ(t)\langle \hat{a}_{1}^{\dag
}\hat{a}_{2}\rangle -iJ(t)\langle \hat{a}_{1}\hat{a}_{2}^{\dag }\rangle -iG_{2}(t)\langle
\hat{a}_{2}^{\dag }\hat{b}\rangle +iG_{2}(t)\langle \hat{a}_{2}\hat{b}^{\dag }\rangle  \notag \\
&&-iG_{2}(t)e^{2i\omega _{m}t}\langle \hat{a}_{2}^{\dag }\hat{b}^{\dag }\rangle
+iG_{2}(t)e^{-2i\omega _{m}t}\langle \hat{a}_{2}\hat{b}\rangle -\kappa
_{2}\langle \hat{a}_{2}^{\dag }\hat{a}_{2}\rangle ,  \notag\\
\frac{d}{dt}\langle \hat{b}^{\dag }\hat{b}\rangle  &=&iG_{2}(t)\langle \hat{a}_{2}^{\dag
}\hat{b}\rangle -iG_{2}(t)\langle \hat{a}_{2}\hat{b}^{\dag }\rangle -i
G_{2}(t)e^{2i\omega _{m}t}\langle \hat{a}_{2}^{\dag }\hat{b}^{\dag }\rangle +i
G_{2}(t)e^{-2i\omega _{m}t}\langle \hat{a}_{2}\hat{b}\rangle  \notag \\
&&-iG_{1}(t)\langle \hat{a}_{1}\hat{b}^{\dag }\rangle +iG_{1}^{\ast }(t)\langle
\hat{a}_{1}^{\dag }\hat{b}\rangle -\gamma _{m}\langle \hat{b}^{\dag }\hat{b}\rangle +\gamma _{m}\bar{%
n}_{m},  \notag \\
\frac{d}{dt}\langle \hat{\hat{a}}_{1}^{\dag }\hat{a}_{2}\rangle  &=&-\left(i\delta +\frac{\kappa
_{1}+\kappa _{2}}{2}\right)\langle \hat{a}_{1}^{\dag }\hat{a}_{2}\rangle -iJ(t)\langle
\hat{a}_{1}^{\dag }\hat{a}_{1}\rangle +iJ(t)\langle \hat{a}_{2}^{\dag }\hat{a}_{2}\rangle  \notag \\
&&-iG_{2}(t)\langle \hat{a}_{1}^{\dag }\hat{b}\rangle -iG_{2}(t)e^{2i\omega
_{m}t}\langle \hat{a}_{1}^{\dag }\hat{b}^{\dag }\rangle +iG_{1}(t)\langle \hat{a}_{2}\hat{b}^{\dag
}\rangle ,  \notag \\
\frac{d}{dt}\langle \hat{a}_{1}^{\dag }\hat{b}\rangle  &=&iJ(t)\langle \hat{a}_{2}^{\dag
}\hat{b}\rangle -iG_{2}(t)\langle \hat{a}_{1}^{\dag }\hat{a}_{2}\rangle -i
G_{2}(t)e^{2i\omega _{m}t}\langle \hat{a}_{1}^{\dag }\hat{a}_{2}^{\dag }\rangle  \notag \\
&&-iG_{1}(t)\langle \hat{a}_{1}^{\dag }\hat{a}_{1}\rangle +iG_{1}(t)\langle \hat{b}^{\dag
}\hat{b}\rangle -\frac{\kappa _{1}+\gamma _{m}}{2}\langle \hat{a}_{1}^{\dag
}\hat{b}\rangle , \notag\\
\frac{d}{dt}\langle \hat{a}_{2}^{\dag }\hat{b}\rangle  &=&\left(i\delta -\frac{\kappa
_{2}+\gamma _{m}}{2}\right)\langle \hat{a}_{2}^{\dag }\hat{b}\rangle +iJ(t)\langle
\hat{a}_{1}^{\dag }\hat{b}\rangle -iG_{2}(t)\langle \hat{a}_{2}^{\dag }\hat{a}_{2}\rangle
+iG_{2}(t)\langle \hat{b}^{\dag }\hat{b}\rangle   \notag \\
&&-i G_{2}(t)e^{2i\omega _{m}t}\langle \hat{a}_{2}^{\dag }\hat{a}_{2}^{\dag
}\rangle +iG_{2}(t)e^{-2i\omega _{m}t}\langle \hat{b}\hat{b}\rangle
-iG_{1}(t)\langle \hat{a}_{1}\hat{a}_{2}^{\dag }\rangle ,  \notag \\
\frac{d}{dt}\langle \hat{a}_{2}^{\dag }\hat{b}^{\dag }\rangle  &=&\left(i\delta -\frac{\kappa
_{2}+\gamma _{m}}{2}\right)\langle \hat{a}_{2}^{\dag }\hat{b}^{\dag }\rangle +iJ(t)\langle
\hat{a}_{1}^{\dag }\hat{b}^{\dag }\rangle +iG_{2}(t)\langle \hat{a}_{2}^{\dag }\hat{a}_{2}^{\dag
}\rangle +iG_{2}(t)\langle \hat{b}^{\dag }\hat{b}^{\dag }\rangle   \notag \\
&&+iG_{2}(t)e^{-2i\omega _{m}t}\langle \hat{a}_{2}^{\dag }\hat{a}_{2}\rangle
+iG_{2}(t)e^{-2i\omega _{m}t}+i G_{2}(t)e^{-2i\omega
_{m}t}\langle \hat{b}^{\dag }\hat{b}\rangle +iG_{1}^{\ast }(t)\langle \hat{a}_{1}^{\dag
}\hat{a}_{2}^{\dag }\rangle ,  \notag \\
\frac{d}{dt}\langle \hat{a}_{1}^{\dag }\hat{b}^{\dag }\rangle  &=&iJ(t)\langle
\hat{a}_{2}^{\dag }\hat{b}^{\dag }\rangle +iG_{2}(t)\langle \hat{a}_{1}^{\dag }\hat{a}_{2}^{\dag
}\rangle +iG_{2}(t)e^{-2i\omega _{m}t}\langle \hat{a}_{1}^{\dag
}\hat{a}_{2}\rangle  \notag \\
&&+iG_{1}(t)\langle \hat{b}^{\dag }\hat{b}^{\dag }\rangle +iG_{1}^{\ast }(t)\langle
\hat{a}_{1}^{\dag }\hat{a}_{1}^{\dag }\rangle -\frac{\kappa _{1}+\gamma _{m}}{2}\langle
\hat{a}_{1}^{\dag }\hat{b}^{\dag }\rangle ,  \notag
\end{eqnarray}
\begin{eqnarray}
\frac{d}{dt}\langle \hat{a}_{1}^{\dag }\hat{a}_{2}^{\dag }\rangle  &=&i\delta \langle
\hat{a}_{1}^{\dag }\hat{a}_{2}^{\dag }\rangle +iJ(t)\langle \hat{a}_{1}^{\dag }\hat{a}_{1}^{\dag
}\rangle +iJ(t)\langle \hat{a}_{2}^{\dag }\hat{a}_{2}^{\dag }\rangle +iG_{2}(t)\langle
\hat{a}_{1}^{\dag }\hat{b}^{\dag }\rangle  \notag \\
&&+iG_{2}(t) e^{-2i\omega _{m}t}\langle \hat{a}_{1}^{\dag }\hat{b}\rangle
+iG_{1}(t)\langle \hat{a}_{2}^{\dag }\hat{b}^{\dag }\rangle -\frac{\kappa
_{1}+\kappa _{2}}{2}\langle \hat{a}_{1}^{\dag }\hat{a}_{2}^{\dag }\rangle ,  \notag \\
\frac{d}{dt}\langle \hat{a}_{1}^{\dag }\hat{a}_{1}^{\dag }\rangle  &=&2iJ(t)\langle
\hat{a}_{1}^{\dag }\hat{a}_{2}^{\dag }\rangle +2iG_{1}(t)\langle \hat{a}_{1}^{\dag }\hat{b}^{\dag
}\rangle -\kappa _{1}\langle \hat{a}_{1}^{\dag }\hat{a}_{1}^{\dag }\rangle , \notag\\
\frac{d}{dt}\langle \hat{a}_{2}^{\dag }\hat{a}_{2}^{\dag }\rangle  &=&(2i\delta-\kappa
_{2} )\langle \hat{a}_{2}^{\dag }\hat{a}_{2}^{\dag }\rangle +2iJ(t)\langle \hat{a}_{1}^{\dag }\hat{a}_{2}^{\dag
}\rangle +2iG_{2}(t)\langle \hat{a}_{2}^{\dag }\hat{b}^{\dag }\rangle +2iG_{2}(t)
e^{-2i\omega _{m}t}\langle \hat{a}_{2}^{\dag }\hat{b}\rangle ,  \notag \\
\frac{d}{dt}\langle \hat{b}^{\dag }\hat{b}^{\dag }\rangle  &=&2iG_{2}(t)\langle
\hat{a}_{2}^{\dag }\hat{b}^{\dag }\rangle +2i G_{2}(t)e^{-2i\omega _{m}t}\langle
\hat{a}_{2}\hat{b}^{\dag }\rangle +2iG_{1}^{\ast }(t)\langle \hat{a}_{1}^{\dag }\hat{b}^{\dag
}\rangle -\gamma _{m}\langle \hat{b}^{\dag }\hat{b}^{\dag }\rangle.
\end{eqnarray}
The equations of motion for other second-order moments can be obtained based on the Hermitian conjugate relations.
\end{widetext}

\end{document}